Banner appropriate to article type will appear here in typeset article

# Water entry of small disks, cones, or anything

**W. Muneo Okiishi[1], Caleb Becker[1], Jesse Belden[2], Aren Hellum[2] and Nathan B. Speirs[1]**

[1]Department of Mechanical Engineering, Brigham Young University, Provo, UT 84602, USA

[2]Naval Undersea Warfare Center, Newport, RI 02841, USA

**Corresponding author:** Nathan Speirs, nspeirs@byu.edu

The water entry of solid and liquid bodies has been studied for over a century, and various researchers have classified the different behaviors that occur when the gas-filled cavity collapses. Although four main cavity collapse regimes have been described and classified for the water entry of small, dense, hydrophobic spheres, only some of these regimes have previously been seen for other impactors, and the scaling used for spheres is not universal across all impactor types. In this paper, we create a unifying scaling to predict cavity collapse regimes, pinch-off time, and pinch-off depth using modified definitions of the Bond, Weber, and Froude numbers for various impactor types. The scaling is based on the downward cavity velocity and an effective diameter, which considers the drag coefficient of the impactor. The impactors we tested include dense solid spheres, disks, and cones, as well as continuous liquid jets and droplet streams. Data for all of these impactor types and behaviors are plotted together with good collapse. Our results indicate that the hydrodynamic characteristics of the impactor, not simply its geometry, govern the global behavior of the cavity.

## 1. Introduction

When an object enters a body of water with sufficient speed, the object pushes the water radially outward, forming a gas-filled cavity in its wake (Worthington & Cole 1900; Richardson 1948; Duclaux *et al.* 2007). For spheres, the impact velocity required to form a cavity varies from just above 0 to ∼ 7 m/s, depending on the static wetting angle (Duez *et al.* 2007) and surface roughness (Zhao *et al.* 2014). As the object descends, the cavity and splash initially expand, but surface tension, a subatmospheric pressure in the cavity, and hydrostatic pressure resist their growth (Aristoff & Bush 2009; Marston *et al.* 2016). Eventually, these forces cause the cavity or the splash to collapse back inward, sealing off further airflow into the cavity.

Water-entry cavities are categorized by the way they collapse, which is commonly called pinch-off. Aristoff & Bush (2009) categorized water-entry cavities created by small, dense, hydrophobic spheres into four regimes based on the depth of the cavity pinch-off, as shown in figure 1. These four regimes occur for differing relative strengths of inertial, surface tension, and gravitational forces. The Bond number $Bo = \rho g d^2 / \sigma$, Weber number $We = \rho U^2 d / \sigma$, and Froude number $Fr = We/Bo = U^2/(gd)$ describe the relative strengths of these forces, where $\rho$ is the liquid density, $g$ is the gravitational acceleration, $d$ is the object diameter, $\sigma$ is the surface tension coefficient, and $U$ is the impact velocity.

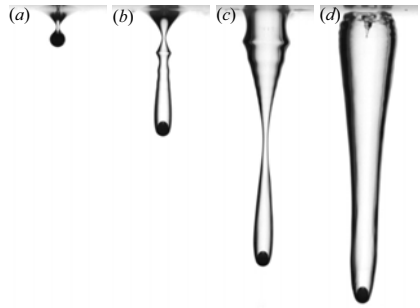

Figure 1. The water entry of dense hydrophobic spheres has been shown to create four main cavity types depending on the Bond and Weber numbers. The cavity types are (*a*) quasi-static seal, (*b*) shallow seal, (*c*) deep seal, and (*d*) surface seal. These images were taken by the authors and previously used in Speirs *et al.* (2019).

At low $We$, the force of surface tension dominates over inertial forces, and a sphere entering the water forms a very small cavity whose walls initially contact the sphere near its equator, forming a shape akin to a static meniscus, as seen in figure 1*a*. As the sphere descends, the contact line slides up the sphere, eventually meeting itself at the apex in a pinch-off type called quasi-static seal.

At higher $We$, spheres form much larger, nearly parabolic cavities in their initial growth phase (Ramsauer & Dobke 1927; May 1975). These cavities collapse either at a shallow depth (on the order of the capillary length (Aristoff & Bush 2009)) due to surface tension, or at a deeper depth (between the middle of the cavity and the sphere, depending on the sphere density (Aristoff *et al.* 2010)) due to hydrostatic pressure, as shown in figure 1*b* & *c*. These collapses are known as shallow seal and deep seal, respectively, and the Bond number $Bo$, which quantifies the relative strength of gravitational (hydrostatic) force compared to surface tension, determines which one occurs.

The final seal type occurs at the highest $We$, when the splash crown collapses atop the open cavity. The splash crown is a thin liquid film with droplets breaking off the top that is ejected around the sphere circumference upon impact. At high $We$, surface tension and the airflow-induced pressure differential across the crown (Marston *et al.* 2016) cause the crown to collapse inward, sealing off further airflow. This collapse is known as surface seal and is shown in figure 1*d*. Each of the four pinch-off regimes occur within a specific region on a $Bo$–$We$ plot, and transition criteria between these regions have been formulated for the water entry of spheres (Aristoff & Bush 2009).

Although these four main cavity collapse regimes are well studied and described for spheres, papers discussing the water entry of other types of impacting bodies do not always identify the collapse regime, even though they see the same behaviors. By applying the cavity classification criteria described by Aristoff & Bush (2009), we can identify the collapse regimes seen for other impacting bodies as well. Quasi-static seal has been observed for small dust particles (Speirs *et al.* 2023). Shallow seal has been observed for multi-droplet streams (Bouwhuis *et al.* 2016). Deep seal has been observed for disks (Bergmann *et al.* 2009), hollow cylinders (Hou *et al.* 2019), high divers (Guillet *et al.* 2020), liquid jets (Kersten *et al.* 2003; Qu *et al.* 2013; Zhu *et al.* 2000), and projectiles with air jets at the front (Zou *et al.* 2024). Both deep and surface seal have been observed for torpedo-like bodies (Bodily *et al.* 2014; Chen *et al.* 2019; Li *et al.* 2020).

Our past studies also show the same cavity collapse regimes and similar $Bo$–$We$ regime plots for different impacting bodies that hint at a unifying scaling to predict collapse behaviors for various impacting bodies. We showed that shallow, deep, and surface seals occur when multi-droplet streams and liquid jets impact a pool (Speirs *et al.* 2018). In a subsequent paper, we showed that the cavity regimes for multi-droplet streams, jets, and spheres can all be predicted with a common $Bo$–$We$ regime plot if the characteristic length scale in $Bo$ and $We$ is changed to the average cavity diameter and the characteristic velocity is changed to the characteristic downward velocity of the cavity (Speirs *et al.* 2019). Although this is interesting, its practical utility is limited, since the average cavity diameter is a dependent parameter and thus difficult to know beforehand for an arbitrary impactor.

In this paper, we develop a unifying scaling to predict the cavity collapse regimes for various impacting bodies on a modified $Bo$–$We$ regime plot regardless of their geometry or phase of matter (solid or liquid) using only independent parameters of the impactor. In the next section, we describe the characteristic length and velocity scales for the modified Bond and Weber numbers. We then describe our experimental study on disk and cone water entry. Finally, in the results section we discuss our data for the disks and cones from this study and the data on spheres, liquid jets, and droplets streams from our previous studies and show common collapse behaviors for various impactor types, discuss our modified $Bo$–$We$ regime plot, describe the regime transitions, and predict the pinch-off time and depth.

## 2. Characteristic length and velocity scales

To find a unifying characteristic length scale for different impactor types, we turn to the adjacent research field of supercavitation. Supercavitation occurs when a submerged body travels at high speed underwater, causing a gaseous cavity to form at the nose (cavitator) of the body that envelops the rest of the body, as sketched in figure 2(*a*). This cavity forms either by water vaporization or air injection and is approximately an axisymmetric ellipsoid with two equal lateral diameters referred to as the cavity diameter $d_c$ and a much larger diameter in the travel direction referred to as the cavity length (Spurk 2002). The cavity diameter is calculated for various nose geometries as $d_c = d_e/\sqrt{Ca}$ using the 'universal linear dimension' or effective diameter $d_e$ of the nose and a non-dimensional cavity number $Ca$ (Spurk 2002; Semenenko 2001). The effective diameter is defined as

$$d_e = d\sqrt{C_D}, \tag{2.1}$$

where $d$ is the nose (cavitator) geometric diameter and $C_D$ is its cavity-running drag coefficient, which is the drag coefficient of the nose with water flowing on the front surfaces and gas on the rear surfaces. The cavity-running drag coefficient has been measured in water tunnel experiments for spheres, disks, and cones of various angles. We derive the drag coefficients for jets and multi-droplet stream impacts in Appendix A. Each of these is listed in Table 1.

For the low velocities in this study, we can simplify the cavity number and consequently the cavity diameter. The cavity number is defined as $Ca = (P_{amb} - P_{cav})/(\rho U^2/2)$, where $P_{amb}$ is the ambient pressure, $P_{cav}$ is the pressure inside the gaseous cavity, $U$ is the travel velocity, and $\rho$ the liquid density (see figure 2(*a*). We follow the suggestion of Aristoff & Bush (2009) and assume that $P_{amb} - P_{cav} \propto \rho_a U^2/2$, where $\rho_a$ is the density of the air flowing into the open cavity. Substituting this into the equation for $Ca$, we find that $Ca$ equals a constant. Consequently, the cavity diameter $d_c \propto d_e$. As described in § 1,

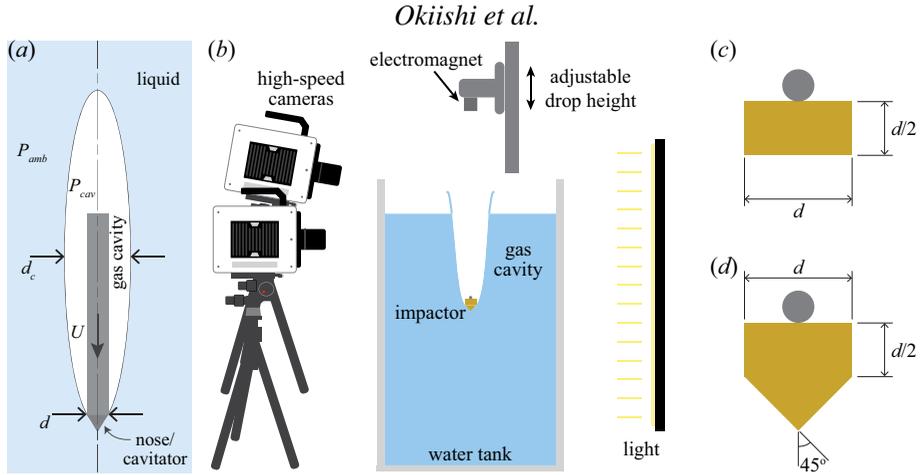

*Okiishi et al.*

Figure 2. (*a*) A schematic of a supercavitating body with an elliptical cavity forming at the nose shows relevant parameters. (*b*) A sketch of the experimental setup used to collect the data for disks and cones is shown. The disk or cone was released from an electromagnet. It then impacted the pool of water while high-speed video was taken both above and below the water surface to visualize the splash and cavity. The geometry is shown for the disks in (*c*) and cones in (*d*).

| Impactor type | $C_D$ | Source |
|---|---|---|
| Sphere | 0.29 | (Hoerner 1965, pg. 10-8) |
| Cone, 45° half angle | 0.48 | (Hoerner 1965, pg. 10-7) |
| Disk | 0.79 | (Hoerner 1965, pg. 10-7) |
| Liquid jet | 4 | Appendix A |
| Droplet stream | 1.31 | Appendix A |

Table 1. The drag coefficients used to calculate $d_e$ for various impactors are listed with their sources.

using the cavity diameter as the characteristic length scale unifies the prediction of cavity regimes for various impacting bodies (Speirs *et al.* 2019). Hence, we expect that we can also use the effective diameter $d_e$ as a unifying characteristic length scale.

The unifying characteristic velocity scale is the characteristic downward velocity of the cavity. Dense solid bodies impacting at relatively low speeds ($U \lesssim 10$ m/s), experience minimal deceleration prior to cavity collapse (Aristoff *et al.* 2010). Hence, the impact velocity is a good approximation of the characteristic velocity. For a jet impacting a pool, the cavity penetrates into the pool at a velocity of $U_c/U_j = [(\rho/\rho_j)^{1/2} + 1]^{-1}$, where $U_j$ is the jet impact velocity, $\rho$ is the pool density, and $\rho_j$ is the jet density (Birkhoff & Zarantonello 1957, p.16). When $\rho_j = \rho$, as in the present case, the cavity velocity simplifies to $U_c/U_j = 1/2$. Hence, for jet impacts, the characteristic velocity is half the impact velocity, $U_j/2$.

The downward velocity of cavities formed by multi-droplet streams is unsteady, increasing as each droplet impacts the bottom of the cavity formed by previous droplets but decreasing between impacts. The average velocity $U_{md}$ increases with the droplet impact frequency, with the upper limit being the same as the velocity of a jet, $U_c = U_{md}/2$ (Bouwhuis *et al.* 2016; Speirs *et al.* 2018). Our previous studies show that the pulsated cavity growth has minimal effect on the cavity dynamics as long as the impact frequency is sufficiently high and that the characteristic velocity for multi-droplet streams is the limit at high frequency, $U_c = U_{md}/2$ (Speirs *et al.* 2018, 2019).

 

Using the effective diameter and the characteristic cavity velocity, we define modified non-dimensional numbers that unify water-entry behaviors for different impactors. The modified Bond number is $Bo_m = \rho g d_e^2/\sigma$. The modified Weber number is $We_m = \rho d_e U_c^2/\sigma$. The modified Froude number is $Fr_m = U_c^2/(g d_e)$.

## 3. Experimental description

The data presented below include results from our previous studies as well as from new experiments on the water entry of disks, cones, and spheres larger than those used in prior work. These previous studies include the entry of hydrophobic ($\theta = 141°$) millimeter-sized steel spheres (Speirs *et al.* 2019) and water streams of multiple droplets and continuous jets (Speirs *et al.* 2018) into a pool of water. We also include the entry of hydrophobic 82 µm solid glass spherical particles into millimetric droplets of water (Speirs *et al.* 2023). The details of the experiments on disks and cones are described next, but the details of the previous studies are described in their respective papers.

Figure 2(*b*) shows the experimental setup we used to examine the water entry of disks and cones in this study. We dropped disks and cones of diameters $d = 2$ to 18 mm from an electromagnet into a pool of water. We adjusted the projectile drop height to vary the impact velocity $U$, which ranged from $U = 0.14$ to 9.65 m/s. Two high-speed cameras recorded the impact event at frame rates of 4,000 frames per second. One of these cameras viewed the impact event above the water surface to image the incoming projectile and splash dynamics. The second camera viewed the impact event below the water surface to image the formation and collapse of the gaseous cavity. From these recordings, we identified the cavity collapse regime of the impact event as defined by Aristoff & Bush (2009) and measured the pinch-off time and the pinch-off depth. For disk impacts, a third camera was often used, which sat below the water level and looked upward at the water surface to image the first contact of the impacting disk and the thin air layer caught between the disk and the pool. This camera recorded up to 200,000 frames per second and enabled us to measure the impact angle of the disks, which was maintained below a few degrees.

We manufactured the disk and cone projectiles from steel with density $\rho_p = 7800$ kg/m³ and brass with density $\rho_p = 8500$ kg/m³. The height of the disks was set equal to the radius $d/2$ as shown in figure 2*c*. The cones had a 45° half angle, and, for ease of manufacturing, the back end of each cone was extended as a cylinder with height also equal to its radius $d/2$, as shown in figure 2*d*. The back ends of each disk and cone had a small indentation in the center, where we adhered a small ferromagnetic sphere so that it would stick to the electromagnet and to improve the alignment of the projectile axis with the vertical direction. For most of the impact events, the disks and cones were uncoated and naturally hydrophilic, but for some trials, the disks and cones were cleaned and then coated with Glaco Mirror Coat Zero, which is a hydrophobic spray that increases the static advancing contact angle to $\theta \approx 140°$ (Speirs *et al.* 2019). We observed no difference between coated and uncoated disks and cones, except at low impact velocities, which we discuss below. We also took additional data on spheres with diameters $d = 29$ and 32 mm to extend the parameter space to larger Bond numbers. These spheres were cleaned and coated with hydrophobic spray to ensure consistent cavity separation.

## 4. Results

Figure 3 shows examples of the cavities formed and a $Bo_m$-$We_m$ plot for spheres, disks, cones, liquid jets, liquid droplet streams, and spherical particles, showing that the effective diameter and characteristic downward cavity velocity unify the water entry regimes for different impactor types. There are some minor differences in the behaviors for different impactor types, which we note below.

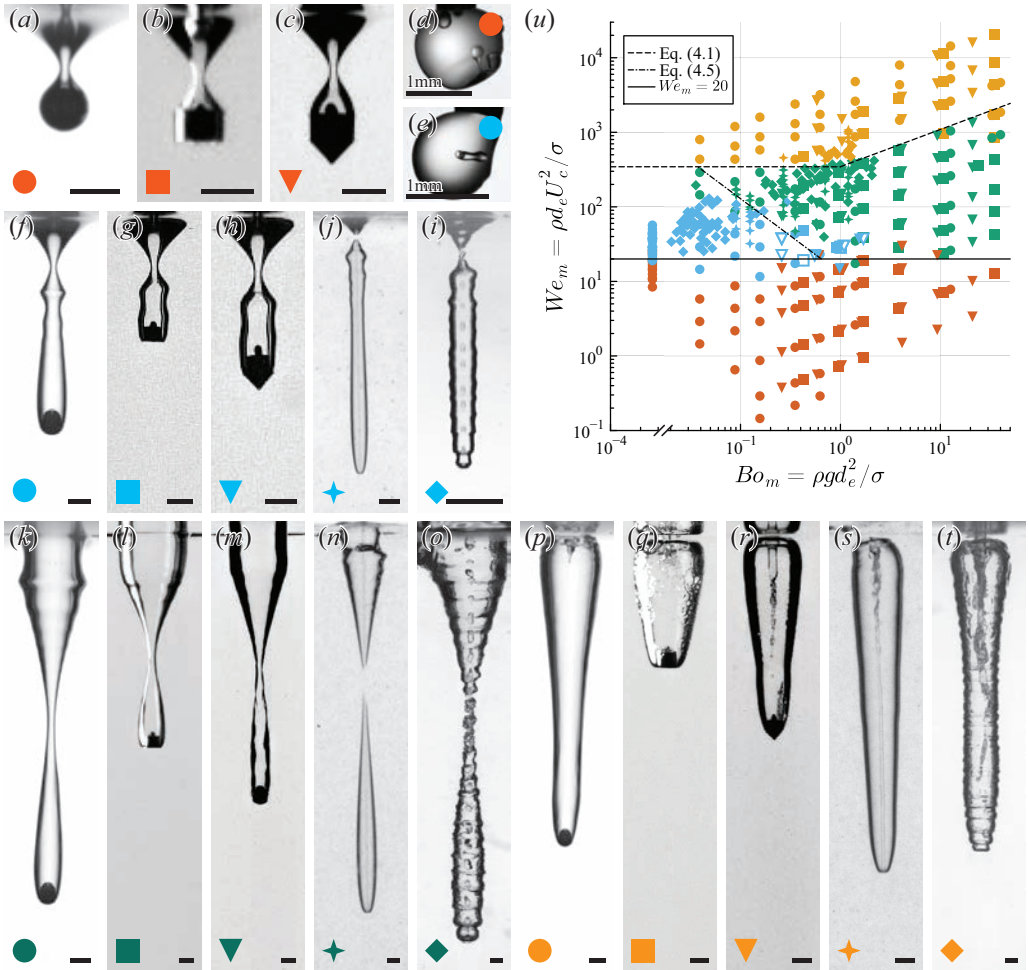

Figure 3. Spheres (circles), disks (squares), cones (triangles), liquid jets (4-pointed stars), and multi-droplet streams (diamonds) experience quasi-static seal (*a–d*), shallow seal (*e–i*), deep seal (*k–o*), and surface seal (*p–t*). The plot in (*u*) shows that the different cavity collapse behaviors occur at common regions on a modified Bond–Weber plot when the effective diameter and cavity velocity are used as the characteristic length and velocity scales. The shapes of each plot symbols indicate the impactor type and their colors indicate the cavity collapse type, as shown on each image (*a-t*). Hollow symbols for shallow seal indicate that the contact line pinned to the bottom corner of the impactor. The lines indicate regime transitions as stated in the legend. The data for the disks, cones, and largest two sphere diameters were taken for this study. Most of the sphere data are from Speirs *et al.* (2019), but the smallest diameter sphere data (particles) are from Speirs *et al.* (2023). The jet and multi-droplet data are from Speirs *et al.* (2018). Some data points have been slightly shifted horizontally to show when repeat trials were conducted and two different regimes were seen. All scale bars are 3 mm wide except the two cases noted otherwise in (*d*) & (*e*). Supplementary Movies 1-4 show full videos of the single images shown in (*a-t*).

At the lowest $We_m$, quasi-static seal only occurs for solid impactors (disks, cones, spheres, and particles), as shown in figure 3(*a–d*) and supplementary movie 1. The liquid impactors combine with the pool upon contact, making it impossible for a quasi-static meniscus to move up their surface and seal at the top, which is the mechanism of quasi-static seal as described for spheres (Aristoff & Bush 2009). For disks and cones in the quasi-static seal regime, we observed two distinct kinds of contact line pinning. On hydrophilic (uncoated) disks and cones, the contact line slides up the cylindrical sides and pins at the sharp upper corner of the cylindrical section as seen in figure 3*b*–*c*. On hydrophobic disks and cones, the contact line usually pins to the lower corner of the cylindrical section, forming a slightly larger cavity. After pinch-off, disks and cones always entrap a bubble on their upper surface due to the contact line pinning at the corners.

We empirically find that once the modified Weber number increases above $We_m \approx 20$, quasi-static seal transitions to shallow or deep seal. This threshold is plotted in figure 3*u* with the solid line and is accurate for nearly six orders of magnitude of $Bo_m$. Within this intermediate $We_m$ range, all six impactor types experience shallow seal at lower $Bo_m$, as seen in figure 3(*e*–*i*) and supplementary movie 2. At higher $Bo_m$, all impactors but particles (which have only small $Bo_m$) experience deep seal, as seen in (*k*–*o*) and supplementary movie 3. For shallow and deep seal, the gaseous cavity becomes much larger and typically separates (i.e. the contact line pins) at the lower corner of disks and cones for hydrophobic bodies, similar to how the cavity separates near the equator for hydrophobic spheres. Hydrophilic disks and cones with $We_m$ near the quasi-static seal regime transition experienced contact-line pinning at both the lower and upper corners, appearing to be in transition between the two regimes. The transition between shallow and deep seal is found by comparing the competing time scales of the two pinch-off types, as both develop simultaneously at different locations. We find this regime transition after discussing the pinch-off times in the next section.

At the highest $We_m$, impactors experience surface seal, as seen in figure 3(*n*–*r*) and supplementary movie 4. The transition to surface seal occurs when the splash crown domes over on top of the cavity and seals off further air ingress before shallow or deep seal occurs. For low Bond number, Aristoff & Bush (2009) empirically found the transition to surface seal to occur at $We = 640$ for spheres. Adjusting this value to account for the effective diameter, the transitional Weber number is $We_m = 640C_{D,sphere}^{1/2} = 345$, which fits the data when $Bo_m \lesssim 1$ (see figure 3(*u*), dashed line). For $Bo_m \gtrsim 1$, we see an increase in the transitional $We_m$ as $Bo_m$ increases and suggest a new empirical fit for the transition to surface seal as follows

$$Bo_m = \begin{cases} 345, & \text{if } Bo \lesssim 1 \\ 345Bo_m^{1/2}, & \text{if } Bo \gtrsim 1. \end{cases} \tag{4.1}$$

### 4.1. *Pinch-off times*

We discuss the time from impact to pinch-off $t_p$ for each collapse behavior using the Weber and Froude numbers. We non-dimensionalize $t_p$ in the usual way, with a velocity and diameter, but whether we use the geometric diameter $d$ or the effective diameter $d_e$ depends on the collapse regime.

For quasi-static seal, we find that the pinch-off time is best predicted with the geometric impactor diameter $d$. This makes sense because large cavities with water at the front of the impactor and air at the back do not form in the quasi-static seal regime. Hence, the flow field around the impactor is different, and the cavity-running drag coefficient is not representative of this regime. As figure 4(*a*) shows, using $d$ in both the non-dimensional pinch-off time $t_pU/d$ and the Froude number $Fr = U^2/(gd)$ results in the data collapsing along a straight line on a semi-log plot. Performing a least squares fit of a logarithmic curve to this data results in

$$\frac{t_pU}{d} = 1.17 \log_{10} Fr + 1.56, \tag{4.2}$$

which is plotted in figure 4(*a*).

Large cavities form for shallow, deep, and surface seals, so the effective diameter $d_e$ can be used to predict their pinch-off times. As shallow seal is caused by surface tension, the modified Weber number $We_m$ best predicts the pinch-off time in this regime, with a curve fit of

$$\frac{t_p U_c}{d_e} = \frac{1}{2} We_m. \quad (4.3)$$

The data and fit are plotted in figure 4(*b*). In some shallow seal impact events for hydrophilic disks and cones, the contact line pins at the upper corner, changing the flow around the projectile and altering $C_D$, the cavity diameter, and thus $d_e$. These data are shown by the hollow symbols in figure 3*u* & 4*b,e* and have been omitted from the above fit.

As deep seal is caused by hydrostatic pressure, the modified Froude number $Fr_m$ best predicts the pinch-off time in this regime with an exponential curve fit of

$$\frac{t_p U_c}{d_e} = 1.82 Fr_m^{1/2}. \quad (4.4)$$

The data and fit are plotted in figure 4(*e*).

We can now use the pinch-off times for shallow and deep seal to find the transition between the two regimes. Equating the pinch-off times for shallow seal (Eq. (4.3)) and deep seal (Eq. (4.4)) and rearranging yields the shallow-to-deep-seal transition

$$We_m = 12.9 Bo_m^{-1}. \quad (4.5)$$

This equation is plotted in figure 3*u* with the dash-dotted line.

We report the pinch-off time for surface seal as the time that the splash crown domes over and contacts itself, sealing off additional air from entering the cavity. As seen in figure 4(*b*), the data for spheres and cones follows a relatively nice trend with $t_p U_c/d_e \approx 50$ at the lowest $We_m$, just after transition to surface seal, which decreases to a constant value of $t_p U_c/d_e \approx 10$ as $We_m$ increases. The data for the disks follows the same pattern, but is shifted downward by a factor of three. This could be due to the difference in splash dynamics very near the body as discussed in Belden *et al.* (2023), however, we leave that exploration to future work. The scatter in the jet and droplet stream data may be explained by the imperfections in the shape of the streams at the leading edge, leading to asymmetric crown formation (Speirs *et al.* 2018).

### 4.2. *Pinch-off depth*

We can predict the depth of the pinch-off location $h_p$ below the free surface for quasi-static, shallow, and deep seal using the Bond, Weber, and Froude numbers. We non-dimensionalize $h_p$ with a diameter, which works best with the geometric diameter $d$ for quasi-static seal and the effective diameter $d_e$ for shallow and deep seals. The pinch-off depth $h_p/d \approx 1$ for quasi-static seal. As $Bo$ increases and hydrostatic pressure becomes more dominant over surface tension, and the pinch-off depth gradually decreases, which is shown by the logarithmic curve fit

$$\frac{h_p}{d_e} = -0.253 \log_{10} Bo + 1.15, \quad (4.6)$$

as plotted in figure 4(*d*).

The pinch-off depth for shallow seal is on the order of the capillary length, as stated by Aristoff & Bush (2009). However, we do see a slight increase in $h_p/d_e$ as $We_m$ increases, although the data is somewhat scattered, as shown in figure 4(*e*). The pinch-off depth for deep seal increases with $Fr_m$ and can be predicted with the following exponential fit, as shown in figure 4(*f*)

$$\frac{h_p}{d_e} = 0.623 Fr_m^{1/2}. \tag{4.7}$$

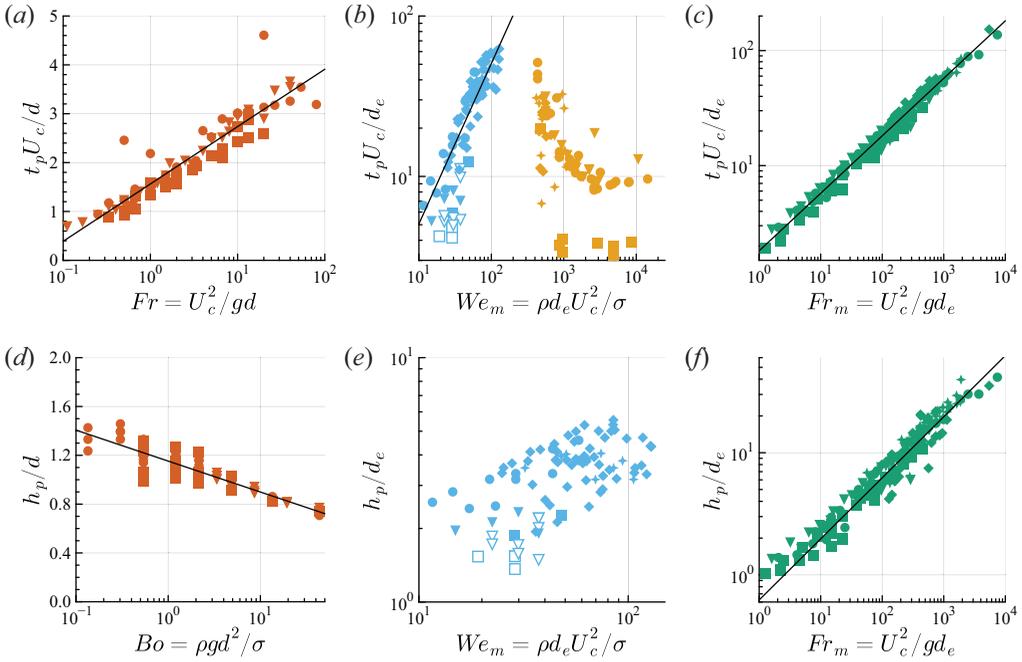

Figure 4. The non-dimensional pinch-off times for quasi-static, shallow, surface, and deep seals are plotted against the appropriate non-dimensional numbers in (*a–c*) respectively. The non-dimensional pinch-off depths for quasi-static, shallow, and deep seals are plotted against the appropriate non-dimensional numbers in (*d–f*) respectively. The marker shapes and colors are the same as those in figure 3. The trend lines in (*a–d,f*) plot Eqs. (4.2)–(4.4) and (4.6)–(4.7) respectively. For shallow seal, hollow symbols indicate that the contact line pinned to the upper corner for disks and cones, while solid symbols indicate that it pinned to the lower corner. The trend line is fit to the solid symbols.

## 5. Conclusion

We see that the proposed scaling unifies the water-entry collapse behavior of various and distinct impactor types, namely solid spheres, disks, and cones and liquid impactors including continuous jets and droplet streams. The unifying characteristic length scale, which is the diameter multiplied by the square root of the drag coefficient, was adopted from the field of supercavitation and is known as the effective diameter. The unifying velocity scale is the characteristic downward velocity of the cavity, which can be approximated as the impact velocity for dense, solid objects and half the impact velocity for liquid jets and droplet streams. Using these characteristic length and velocity scales to calculate the Bond, Weber, and Froude numbers enables a unified prediction of the cavity collapse regimes, pinch-off times, and pinch-off depths for all impactor types with only minor deviations. Future research could seek to extend this unified scaling to solid bodies with density similar to or less than the density of the pool fluid, as significant deceleration will occur prior to cavity collapse, altering collapse behaviors.

**Funding information**  W.M.O., C.B., and N.B.S. acknowledge funding from the BYU Ira A. Fulton College of Engineering. J.B. and A.M.H acknowledge funding from the Naval Undersea Warfare Center In-House Laboratory Independent Research program, monitored by Dr. Elizabeth Magliula.

**Declaration of interests**  The authors report no conflict of interest.

REFERENCES

ARISTOFF, JEFFERY M. & BUSH, JOHN W. M. 2009 Water entry of small hydrophobic spheres. *Journal of Fluid Mechanics* **619**, 45–78.

ARISTOFF, JEFFREY M., TRUSCOTT, TADD T., TECHET, ALEXANDRA H. & BUSH, JOHN W. M. 2010 The water entry of decelerating spheres. *Physics of Fluids* **22** (3), 032102.

BELDEN, JESSE, SPEIRS, NATHAN B., HELLUM, AREN M., JONES, MATTHEW, PAOLERO, ANTHONY J. & TRUSCOTT, TADD T. 2023 Water entry of cups and disks. *Journal of Fluid Mechanics* **963**, A14.

BERGMANN, RAYMOND, VAN DER MEER, DEVARAJ, GEKLE, STEPHAN, VAN DER BOS, ARJAN & LOHSE, DETLEF 2009 Controlled impact of a disk on a water surface: cavity dynamics. *Journal of Fluid Mechanics* **633**, 381–409.

BIRKHOFF, GARRETT & ZARANTONELLO, E. H. 1957 *Jets, Wakes, and Cavities (Series in Applied Mathematics and Mechanics)*, , vol. 2. Academic Press.

BODILY, KYLE G., CARLSON, STEPHEN J. & TRUSCOTT, TADD T. 2014 The water entry of slender axisymmetric bodies. *Physics of Fluids* **26** (7), 072108.

BOUWHUIS, WILCO, HUANG, XIN, CHAN, CHON U, FROMMHOLD, PHILIPP E., OHL, CLAUS-DIETER, LOHSE, DETLEF, SNOEIJER, JACCO H. & VAN DER MEER, DEVARAJ 2016 Impact of a high-speed train of microdrops on a liquid pool. *Journal of Fluid Mechanics* **792**, 850–868.

CHEN, CHEN, SUN, TIEZHI, WEI, YINGJIE & WANG, CONG 2019 Computational analysis of compressibility effects on cavity dynamics in high-speed water-entry. *International Journal of Naval Architecture and Ocean Engineering* **11** (1), 495–509.

DUCLAUX, V., CAILLÉ, F., DUEZ, C., YBERT, C., BOCQUET, L. & CLANET, C. 2007 Dynamics of transient cavities. *Journal of Fluid Mechanics* **591**, 1–19.

DUEZ, CYRIL, YBERT, CHRISTOPHE, CLANET, CHRISTOPHE & BOCQUET, LYDÉRIC 2007 Making a splash with water repellency. *Nature Physics* **3** (3), 180–183.

GUILLET, THIBAULT, MOUCHET, MÉLANIE, BELAYACHI, JÉRÉMY, FAY, SARAH, COLTURI, DAVID, LUNDSTAM, PER, HOSOI, PEKO, CLANET, CHRISTOPHE & COHEN, CAROLINE 2020 The hydrodynamics of high diving. *Proceedings* **49** (1).

HOERNER, SIGHARD F. 1965 *Fluid-Dynamic Drag*. Published by the author.

HOU, YU, HUANG, ZHENGUI, CHEN, ZHIHUA, GUO, ZEQING & LUO, YUCHUAN 2019 Investigations on the vertical water-entry of a hollow cylinder with deep-closure pattern. *Ocean Engineering* **190**, 106426.

KERSTEN, B., OHL, C. D. & PROSPERETTI, A. 2003 Transient impact of a liquid column on a miscible liquid surface. *Physics of Fluids* **15** (3), 821–824, arXiv: https://pubs.aip.org/aip/pof/article-pdf/15/3/821/19221129/821_1_online.pdf.

LI, ZHIPENG, SUN, LONGQUAN, YAO, XIONGLIANG, WANG, DULIANG & LI, FOCHEN 2020 Experimental study on cavity dynamics in high froude number water entry for different nosed projectiles. *Applied Ocean Research* **102**, 102305.

MARSTON, J. O., TRUSCOTT, T. T., SPEIRS, N. B., MANSOOR, M. M. & THORODDSEN, S. T. 2016 Crown sealing and buckling instability during water entry of spheres. *Journal of Fluid Mechanics* **794**, 506–529.

MAY, A. 1975 Water entry and the cavity-running behavior of missiles. *Tech. Rep.* SEAHAC/TR 75-2. Naval Surface Warfare Center White Oaks Laboratory.

QU, XIAOLIANG, GOHARZADEH, AFSHIN, KHEZZAR, LYES & MOLKI, ARMAN 2013 Experimental characterization of air-entrainment in a plunging jet. *Experimental Thermal and Fluid Science* **44**, 51–61.

RAMSAUER, CARL & DOBKE, G. 1927 Die bewegungserscheinungen des wassers beim durchgang schnell bewegter kugeln. *Annalen der Physik* **389** (22), 697–720.

RICHARDSON, E. G. 1948 The impact of a solid on a liquid surface. *Proc. Phys. Soc.* **4**, 352–367.

SEMENENKO, VLADIMIR N. 2001 Artificial supercavitation. physics and calculation. *Tech. Rep.*. Ukrainian National Academy of Sciences - Institute of Hydromechanics.

SPEIRS, NATHAN B., BELDEN, JESSE L. & HELLUM, AREN M. 2023 The capture of airborne particulates by rain. *Journal of Fluid Mechanics* **958**, A40.

SPEIRS, NATHAN B., MANSOOR, MOHAMMAD M., BELDEN, JESSE & TRUSCOTT, TADD T. 2019 Water entry of spheres with various contact angles. *Journal of Fluid Mechanics* **862**, R3.

SPEIRS, NATHAN B., PAN, ZHAO, BELDEN, JESSE & TRUSCOTT, TADD T. 2018 The water entry of multi-droplet streams and jets. *Journal of Fluid Mechanics* **844**, 1084–1111.

Rapids articles must not exceed this page length

Spurk, J. H. 2002 On the gas loss from ventilated supercavities. *Acta Mechanica* **155** (3), 125–135.

Worthington, Arthur Mason & Cole, R. S. 1900 Iv. impact with a liquid surface studied by the aid of instantaneous photography. paper ii. *Philosophical Transactions of the Royal Society of London. Series A, Containing Papers of a Mathematical or Physical Character* **194** (252-261), 175–199.

Zhao, Meng-Hua, Chen, Xiao-Peng & Wang, Qing 2014 Wetting failure of hydrophilic surfaces promoted by surface roughness. *Scientific Reports* **4** (1), 5376.

Zhu, Yonggang, Oguz, Hasan N. & Prosperetti, Andrea 2000 On the mechanism of air entrainment by liquid jets at a free surface. *Journal of Fluid Mechanics* **404**, 151–177.

Zou, Zhihui, Jiang, Yunhua & Wu, Bin 2024 Cavity dynamics of vertical water entry by air jet. *Phys. Rev. Fluids* **9**, 124006.

## Appendix A. Drag coefficients for jets and multi-droplet streams

To calculate $d_e$ for liquid jets and multi-droplet streams, we first need their cavity running drag coefficient $C_D$, which we now derive for a jet and then adapt for a multi-droplet stream. When a liquid jet impacts a liquid pool, the jet pushes the pool outward, forming a gas-filled cavity that is coated with the jet fluid (Zhu *et al.* 2000; Kersten *et al.* 2003). If the jet liquid and pool liquid are the same then the cavity velocity $U_c$ over the jet velocity $U_j$ is $U_c/U_j = 1/2$ (Birkhoff & Zarantonello 1957, p.16). Using the integral conservation of momentum on a control volume enclosing the jet fluid and moving steadily downward at velocity $U_c$, as shown in supplementary figure 5, we can write

$$\vec{F_D} = \int_{CS1} \vec{W}_1 \rho (\vec{W}_1 \cdot \vec{n_1}) dA + \int_{CS2} \vec{W}_2 \rho (\vec{W}_2 \cdot \vec{n_2}) dA \tag{A 1}$$

where $\vec{W}_1$, $\vec{W}_2$, $\vec{n_1}$, and $\vec{n_2}$ are the relative velocities and outward normals at control surfaces (CS) 1 and 2 respectively. Notice that we have neglected viscous and gravitational forces as the Reynolds and Froude numbers are much greater than 1. By applying the Bernoulli equation between a point on CS1 and a point on CS2, and noting that the pressure and height between these points is approximately equal, $P_1 \approx P_2$ and $z_1 \approx z_2$, we see that the velocity into CS1 is approximately equal to the velocity out of CS2, $U_1 \approx U_2$. By conservation of mass we also see that the area that the jet fluid passes through to enter the control volume and exit the control volume are also approximately equal, $A_2 \approx A_1 = A_j$, where $A_j$ is the cross-sectional area of the jet. Plugging each of these terms into Eq. (A 1) and simplifying yields the drag force in the upward direction,

$$F_D = \frac{1}{2} \rho U_j^2 A_j. \tag{A 2}$$

If we now insert the typical equation for the drag force and define the jet drag coefficient in terms of the cavity penetration velocity $\frac{1}{2} U_j$ and jet area $A_j$ as follows

$$\frac{1}{2} \rho \left( \frac{1}{2} U_j \right)^2 A_j C_{D,j} = \frac{1}{2} \rho U_j^2 A_j \tag{A 3}$$

we get that $C_{D,j} = 4$ for a liquid jet that forms a cavity while impacting a pool of the same fluid.

To find the cavity-running drag coefficient for a multi-droplet stream, we look at our previous work (Speirs *et al.* 2018). In that work, we found that a multi-droplet stream and a jet impacting a pool at the same velocity create the same cavity diameter when the droplet diameters $d_{md}$ are 1.75 times greater than the jet diameter $d_j$. Although we did not have a theoretical explanation for this at the time, we now believe that this is due to the difference in the cavity-running drag coefficients. If the cavity diameters are equal for the same impact velocity, then their effective diameters must also be equal, $d_{e,j} = d_{e,md}$. By Eq. (2.1), this means that $d_j \sqrt{C_{D,j}} = d_{md} \sqrt{C_{D,md}}$. Substituting $d_j = d_{md}/1.75$ and rearranging we find that the cavity-running drag coefficient for a multi-droplet stream based on the droplet diameter and characteristic velocity of half the impact velocity $\frac{1}{2} U_{md}$ is $C_{D,md} = C_{D,j}/1.75^2 \approx 1.31$.

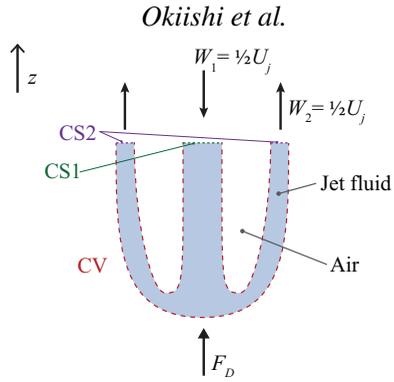

Supplementary Figure 5. A sketch showing the control volume (CV), control surfaces (CS1 & CS2), flow velocities, and drag force use in the integral conservation of momentum equation to find the cavity-running drag coefficient for a liquid jet impacting a liquid pool. The control volume moves downward with the velocity of the bottom of the cavity $1/2U_j$.